\documentclass[sn-aps]{sn-jnl}

\usepackage{graphicx}%
\usepackage{multirow}%
\usepackage{amsmath,amssymb,amsfonts}%
\usepackage{amsthm}%
\usepackage{mathrsfs}%
\usepackage[title]{appendix}%
\usepackage{xcolor}%
\usepackage{textcomp}%
\usepackage{manyfoot}%
\usepackage{booktabs}%
\usepackage{algorithm}%
\usepackage{algorithmicx}%
\usepackage{algpseudocode}%
\usepackage{listings}%
\usepackage{orcidlink}%

\begin{document}

\title[A Note on Norton's Dome]{A Note on Norton's Dome}


\author*[1]{\fnm{Christine} \sur{C. Dantas \,\orcidlink{0000-0002-2833-2520}}}\email{christine.dantas@inpe.br}

\affil*[1]{\orgdiv{christine.dantas@inpe.br}, \orgname{Instituto Nacional de Pesquisas Espaciais}, \orgaddress{\street{Av. dos Astronautas, 1758, Jardim da Granja}, \city{S\~ao Jos\'e dos Campos}, \postcode{12227-010}, \state{SP}, \country{Brazil}}}


\abstract{``Norton's Dome'' is an example of a Newtonian system that violates the Lipschitz condition at a single point, leading to non-unique solutions (indeterminism). Here we reformulate this problem into a ``weak'' form (in the sense of distributions). In our description the indeterminism manifests through the problematic interpretation of initial conditions, since distributions (as linear functionals on the space of test functions) do not have values at individual points. }


\maketitle

\section{Introduction} 
The so-called ``Norton's Dome'' problem (NDP) admits a family of solutions describing the spontaneous motion of a particle, initially at rest on the top of a dome in a gravitational field \cite{Nor03}, \cite{Nor08}. This interesting problem is considered an example of indeterminism in Newtonian mechanics and some critical accounts have been published in the past years (e.g., \cite{Kor06}, \cite{Kos08}, \cite{Mal08}, \cite{Wil09}, \cite{Rob09}, \cite{Fle12}, \cite{Lar13}). From the mathematical point of view, the indeterminism arises in this system due to a violation of the Lipschitz  condition (c.f. \cite{Fle12} and references therein).

In this note, we show that the NDP can be described in ``weak'' (distributional) form. Our motivation is to present a simple and mathematically consistent alternative to the original problem, perhaps raising additional philosophical issues regarding the nature of indeterminism in such systems.

The NDP is described by a radially symmetric dome with the following shape (see details in the original papers by Norton, \cite{Nor03}, \cite{Nor08}):
\begin{equation}
h = {2 \over 3} g r^{3 \over 2},
\end{equation}
\noindent where $h$ is the height of the dome's top (set at $r = 0$, where $r$ is the radial distance coordinate on the frictionless surface of the dome), with a unit mass particle placed at rest on the top, at time $t = 0$. The particle is subject to a gravitational field, directed downward, with acceleration due to gravity $g$. It can be shown that the net force (the component
of the gravitational force tangent to the surface) on the particle is:
\begin{equation}
F = g {dh \over dr} = r^{1 \over 2}.
\end{equation}

Note that the gravitational force can only
accelerate the particle along the surface. Newton's second law states that this force is equal to the particle's acceleration, $a(t) = d^2r/dt^2$. This gives the particle's  equation of motion:
\begin{equation}
{d^2 r \over dt^2} = r^{1 \over 2},  \label{EoM}
\end{equation}
\noindent subject to the initial conditions:
\begin{equation}
r(t=0) = 0,  \hspace{2cm} {d r \over dt}\Bigr|_{\substack{t=0}} = 0. \label{IC}
\end{equation}

Clearly, the solution in which the particle stays at rest on the top, for all times, is the physically expected (trivial) solution:
\begin{equation}
r(t) = 0, \hspace{2cm} \forall ~ t. \label{trivial}
\end{equation}

However, Norton points out \cite{Nor03}, \cite{Nor08} that there is also a family of admissible solutions, for arbitrary radial direction and arbitrary time $T$, given by:
\begin{equation}
r(t) =
\left\{ \begin{array}{lcl}
(1/144) (t-T)^4, & \mbox{for}
& t \ge T, \\ 
0, & \mbox{for} & t \le T.
\end{array}\right.  \label{Norton_r}
\end{equation}
\noindent Taking the second derivative with respect to time to Norton's proposed solution above (Eq. \ref{Norton_r}), one obtains:
\begin{equation}
{d^2 r \over dt^2} = a(t) =
\left\{ \begin{array}{lcl}
(1/12) (t-T)^2, & \mbox{for}
& t \ge T, \\ 
0, & \mbox{for} & t \le T.
\end{array}\right.  \label{Norton_a}
\end{equation}
\noindent By noting that Eq. \ref{Norton_a} is the square root of Eq. \ref{Norton_r}, Norton concludes that, indeed,  Eq. \ref{Norton_r} is an admissible solution to the particle's equation of motion, Eq. \ref{EoM}, with the initial conditions given in Eq. \ref{IC}. Note that in this non-trivial solution, the particle spontaneously starts moving at an arbitrary time $T$.  

We point out that we have written Eqs. \ref{Norton_r} and \ref{Norton_a} exactly as stated by Norton \cite{Nor03} \cite{Nor08}, namely, with the inequalities $\ge$ and $\le$ in both conditions relating $t$ and $T$. This is specially confusing, given the importance of understanding the meaning of the exact time $t = T$. Norton discusses this ``crucial'' time at some length; particularly,  he states \cite{Nor03}:
\begin{quote}
\normalsize{
We are tempted to think of the instant $t=T$ as the first instant at which the mass moves. But that is not so. It is the {\it last} instant at which the mass does {\it not} move. There is no first instant at which the mass moves.}
\end{quote}

In our exposition, we do not discuss the ``crucial time'' $t=T$ along the lines given by Norton. 
We will proceed by pointing out that the NDP can be reformulated in terms of distributions, so that the emphasis on the  ``crucial time'' is diminished. 

\section{How the NDP can be reformulated in terms of distributions}

We begin by re-writing Norton's solution, Eq. \ref{Norton_r}, as a single expression. Let $x \equiv (t-T)$, then:
\begin{equation}
r(x) = {1 \over 144} x^4 \Theta(x),  \label{WeakSol}
\end{equation}
\noindent where $\Theta$ is the Heaviside (step) function:
\begin{equation}
\Theta(x) = 
\left\{ \begin{array}{lcl}
1, & \mbox{for}
& x \ge 0, \\ 
0, & \mbox{for} & x < 0.
\end{array}\right.  \label{Step}
\end{equation}
\noindent This re-writing does not differ from Norton's solution, Eq. \ref{Norton_r}, except for our right-continuous definition of the step function, setting $\Theta(0) = 1$. As mentioned previously, Eq. \ref{Norton_r} is somewhat confusing regarding the condition $t = T$, requiring additional clarifications (some of which are given by Norton in \cite{Nor03}). It is not clear to us how the paths at times $ t \le T$ and $t \ge T$ are supposed to be connected at $t = T$, either in the mathematical or in the physical sense.  By considering Eq. \ref{WeakSol} instead, we bypass such a potential ambiguity. The choice of the zero argument of the step function is immaterial by regarding it as a distribution. Furthermore, we have the mathematical apparatus of analysis of distributions, which will guide us into a more rigorous exposition (e.g., \cite{Kol75}).

Consider the space $C_0^{\infty}(\mathbb{R})$ of smooth, compactly supported, test functions $\phi$, defined on $\mathbb{R}$ (see, e.g., Tao's notes in \cite{TaoBlog}). A distribution is defined in terms of a continuous, linear functional, $T_f:C_0^{\infty}(\mathbb{R}) \rightarrow \mathbb{R}$. The denote the associated space of distributions on $\phi$ as $D^{\prime}$. An explicit construction of such a linear functional is \cite{Kol75}:
\begin{equation}
T_f(\phi) = (f,\phi) = \int_{-\infty}^{\infty} f(x) \phi(x) dx, \label{LF}
\end{equation}
\noindent where $f$ is a given locally integrable function. In other words, for each $\phi$, we associate a number, given above, Eq. \ref{LF}, involving the function $f$\footnote{For example, the Dirac delta function, $\delta(x) = 1 ~({\rm for}~ x = 0) ~{\rm and} ~ \delta(x) = 0 ~({\rm for}~ x \ne 0)$, is not actually a function in the usual sense, but a linear functional on a space of test functions, defined as:
\begin{equation}
T_{\delta}(\phi) = (\delta,\phi) = \int_{-\infty}^{\infty} \delta(x) \phi(x) dx = \phi(0). \label{DiracDelta}
\end{equation}
}. For example, setting $f(x) = \Theta(x)$ in Eq. \ref{LF}, a linear functional associated with the step function would be \cite{Kol75}:
\begin{equation}
T_{\Theta}(\phi) = (\Theta,\phi) = \int_{-\infty}^{\infty} \Theta(x) \phi(x) dx = 
\int_{0}^{\infty} \phi(x) dx . \label{StepLinearFunctional}
\end{equation}

In the theory of distributions, the common notion of differentiation is basically transfered to 
test functions, where they are well defined; hence we are able to make sense of the ``derivative'' of the highly ``singular functions''. The {\it weak} or {\it distributional derivative} $\mathcal{D}$ of a distribution can be shown to obey \cite{Kol75}:
\begin{equation}
\mathcal{D} [T_f(\phi)] = - \int_{-\infty}^{\infty} f(x) \phi^{\prime} (x) dx = -T_f(\phi^{\prime}), \label{derivaT}
\end{equation}
\noindent where the prime symbol refers to the usual derivative with respect to the independent variable. Higher ``weak'' derivatives are obtained in the same manner, giving:
\begin{equation}
\mathcal{D}^n [T_f(\phi)] = (-1)^n \int_{-\infty}^{\infty} f(x) \phi^{(n\prime)} (x) dx . \label{derivaTn}
\end{equation}

Our task here is find linear functionals, $T(\phi)$, associated with the functions appearing in the equation of motion, Eq. \ref{EoM}, namely, $f_1(x) = r(x)$, and $f_2(x)  = \sqrt{r(x)}$, in order to obtain a ``weak'' condition, where derivatives in the equation of motion are distributional derivatives. A ``weak'' version of the NDP would, then, be:
\begin{equation}
\mathcal{D}^2 [T_{f_1}(\phi)] = T_{f_2}(\phi). \label{WeakNDP}
\end{equation}

The most trivial and direct manner to satisfy the condition above is to find appropriate linear functionals leading to identically zero values on both sides. This choice is also conceptually interesting because it resembles the trivial solution of the original NDP, Eq. \ref{trivial}.  Since Eq. \ref{WeakSol} involves a polynomial, we must find suitable test functions, given that a polynomial go to infinity as $| x | \rightarrow \infty $, therefore test functions must go to zero as $| x | \rightarrow \infty $ faster than any inverse power of $x$. We are able to find such linear functionals by moving to the larger space of {\it Schwartz test functions}, $\mathcal{S}(\mathbb{R}) \supset C_0^{\infty}(\mathbb{R}) $, on which the (more restricted) {\it tempered distributions} (belonging to the space $S^{\prime} \subset D^{\prime}$) act upon \cite{TaoBlog}. 

We select the following set of test functions, for reasons which will become clear below:
\begin{equation}
\phi(x)_k = P_k(x) \exp(-x^4), \label{TestF}
\end{equation}
\noindent where $P_k(x)$ is some polynomial (labeled by $k$). Since any polynomial multiplied by a Schwartz test function, $\exp(-x^4)$, is also a Schwartz test function, our choice above is acceptable.
Then suitable linear functionals for the ``weak'' NDP, based on Eq. \ref{WeakSol}, would be:
\begin{equation}
T_{f_1}(\phi) \equiv \int_{-\infty}^{\infty} r(x) \phi_1(x)  dx = 
\int_{-\infty}^{\infty}\Theta(x)  \bar{\phi}_1(x)  dx = 
\int_{0}^{\infty} \bar{\phi}_1(x) dx = T_{\Theta}(\bar{\phi}_1),  \label{Tr}
\end{equation}
\noindent and
\begin{equation}
T_{f_2}(\phi) \equiv \int_{-\infty}^{\infty} \sqrt{r(x)} \phi_2(x)  dx = 
\int_{-\infty}^{\infty}\Theta(x)  \bar{\phi}_2(x)  dx = 
\int_{0}^{\infty} \bar{\phi}_2(x) dx = T_{\Theta}(\bar{\phi}_2), \label{Tsqrtr}
\end{equation}
\noindent with
\begin{equation}
\bar{\phi}_1(x) = \left ( {x^4 \over 144} \right )  P_1(x) \exp(-x^4), \label{Phi1}
\end{equation}
\and 
\begin{equation}
\bar{\phi}_2(x) = \left ( {x^2 \over 12} \right )   P_2(x) \exp(-x^4), \label{Phi2}
\end{equation}
\noindent where we have used properties of the step functional given by Eq. \ref{StepLinearFunctional} and the fact that $\Theta(x) = \sqrt{\Theta(x)}$. By applying the rules of distributional derivatives, Eq. \ref{derivaTn}, twice on Eq. \ref{Tr}, we have:
\begin{equation}
\mathcal{D}^2 [T_{\Theta}(\bar{\phi}_1)] =  \int_{0}^{\infty} \bar{\phi}_1^{\prime\prime} (x) dx . \label{derT2}
\end{equation}
\noindent After a few attempts, the simplest choice, $P_1(x) = 1$, inserted into Eq. \ref{Phi1}, and applied above, leads to:
\begin{equation}
\mathcal{D}^2 [T_{\Theta}(\bar{\phi}_1)] =  { 1 \over 36}
\int_{0}^{\infty}  \exp({-x^4}) x^2 (4 x^8 - 11 x^4 + 3) dx = 0. \label{derT2B}
\end{equation}
\noindent It is not difficult to find a polynomial $P_2(x)$ that leads to a similar integral giving zero. The result above indicates that choosing
\begin{equation}
P_2(x) = 4 x^8 - 11 x^4 + 3 ,
\end{equation}
\noindent and applying into Eqs. \ref{Phi2}, \ref{Tsqrtr}, also leads to $T_{\Theta}(\bar{\phi}_2) = 0 $. This simple exercise shows that the ``weak'' version of the NDP, Eq. \ref{WeakNDP}, can be satisfied trivially by the linear functionals chosen above.

As a counter-example, the choice, e.g., $P_2(x) = 1$, does not lead to a zero integral, i. e.,

\begin{equation}
T_{f_2}(\phi) = T_{\Theta}(\bar{\phi}_2) = \int_0^{\infty} \left ( {x^2 \over 12} \right ) \exp(-x^4) = 
{\Gamma \left ( {3 \over 4} \right ) \over 48 },
\end{equation}
\noindent where $\Gamma(x)$ is the Gamma function.

\section{Conclusions}

In this note we did not explore the possible conceptual reinterpretations of our ``weak'' formulation, leaving them open for future works. In terms of the original formulation, the explanation that the NDP violates the Lipschitz condition at a single point (the top), leading to non-unique solutions \cite{Kor06}, \cite{Fle12}, conceptually clarifies various philosophical issues regarding indeterminism. Possible conceptual relations between this violation and the ``weak'' formulation are left open.

Although we have not extensively searched for other admissible linear functionals satisfying  Eq. \ref{WeakNDP}, it is possible that a general proof can be found in a straightforward way, but we did not address this issue here. Our motivation was merely to reformulate the NDP in terms of distributions in a clear manner, and to show that it can be made mathematically consistent in this way. By this, the extreme idealization of the original NDP, along with its non-Lipschitz characteristics, are not issues in the ``weak'' form, furnishing a more physically satisfying description, specially since it allows us to introduce probability measures, as shown in Ref. \cite{GenFlows03}, for problems involving non-Lipschitz flows. 

In other words, the ``weak'' NDP allows for the inclusion of small initial perturbations modeled from a probability distribution, leading to acceptable solutions (in the ``weak'' sense), in which the particle does move from the top at some time $t = T$. This time would also seem to be arbitrary for an arbitrary choice of the probability distribution but, on the other hand, it would {\it not} seem to be arbitrary for a probability distribution coherent with respect to the physical conditions prevailing in a much more ``realistic'' dome.

Finally, we point out that the initial value conditions cannot be interpreted in a straighforward manner for the ``weak'' NDP. We quote from this note \cite{DeltaNotes}:

\begin{quote}
\normalsize{
Unfortunately, distributions are not a free lunch; they come with their own headaches. (...) Since distributions do not have values at individual points, it is not so easy to impose boundary conditions on the solutions if they are viewed as distributions--- what does it mean to set $u(0) = 0$? There are ways around this, but they are a bit cumbersome, especially in more than one dimension.
}
\end{quote}

Therefore, in the ``weak'' description of the NDP, the indeterminism in the original problem manifests through the problematic intepretation of initial conditions in distributions. They should be conceptually transfered to the notion of probability measures. We leave open related philosophical issues, specially if there is a conceptual relation between both views (the original and the ``weak'' NDPs), in terms of what constitutes the fundamental (not only operational) conditions for establishing valid Newtonian systems, if there are any such conditions at all.

\bmhead{Acknowledgements}

This study was financed in part by the Coordena\c c\~ao de Aperfei\c coamento de Pessoal de N\'{\i}vel Superior - Brasil (CAPES) - Finance Code 001, and the Brazilian Space Agency (AEB) for the funding (PO 20VB.0009).


\end{document}